# Fairy Lights in Femtoseconds: Aerial and Volumetric Graphics Rendered by Focused Femtosecond Laser Combined with Computational Holographic Fields

Yoichi Ochiai[1]*   Kota Kumagai[2]*   Takayuki Hoshi[3]   Jun Rekimoto[4]   Satoshi Hasegawa[2]   Yoshio Hayasaki[2]

[1]University of Tsukuba   [2]Utsunomiya University   [3]Nagoya Institute of Technology   [4]The University of Tokyo

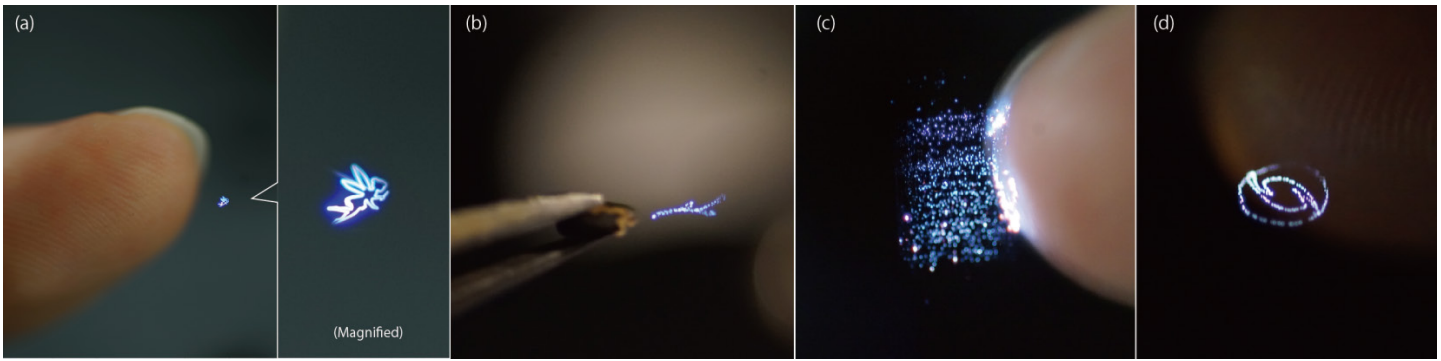

**Figure 1.** *Application images of Fairy Lights in Femtoseconds, aerial and volumetric graphics in air rendered by femtosecond lasers. (a) A "fairy" flying in front of a finger. (b) A "sprout" coming out from a seed. (c) Interference between a point cloud and a finger. (d) The SIGGRAPH logo.*

## Abstract

We present a method of rendering aerial and volumetric graphics using femtosecond lasers. A high-intensity laser excites a physical matter to emit light at an arbitrary 3D position. Popular applications can then be explored especially since plasma induced by a femtosecond laser is safer than that generated by a nanosecond laser. There are two methods of rendering graphics with a femtosecond laser in air: Producing holograms using spatial light modulation technology, and scanning of a laser beam by a galvano mirror. The holograms and workspace of the system proposed here occupy a volume of up to 1 cm³; however, this size is scalable depending on the optical devices and their setup. This paper provides details of the principles, system setup, and experimental evaluation, and discussions on scalability, design space, and applications of this system. We tested two laser sources: an adjustable (30-100 fs) laser which projects up to 1,000 pulses per second at energy up to 7 mJ per pulse, and a 269-fs laser which projects up to 200,000 pulses per second at an energy up to 50 μJ per pulse. We confirmed that the spatiotemporal resolution of volumetric displays, implemented with these laser sources, is 4,000 and 200,000 dots per second. Although we focus on laser-induced plasma in air, the discussion presented here is also applicable to other rendering principles such as fluorescence and microbubble in solid/liquid materials.

**CR Categories:** H.5 [Information interfaces and presentation];

**Keywords:** Volumetric display, Laser plasma, Femtosecond laser, Aerial interaction, Touchable aerial images

---

*\* Joint first authors*

## 1 Introduction

Three-dimensional (3D) displays have attracted great attention over the past five decades. 3D virtual objects were originally displayed with a head-mounted display in [Sutherland 1968]. Since then, continuous efforts have been made to explore 3D displays that have planar surfaces, and several methods have been developed to provide stereopsis for binocular vision [Benzie et al. 2007]. The technologies that employ glasses to achieve this are based on such as anaglyphs, time-division, and polarization. On the other hand, those technologies that do not rely on glasses are based on such as parallax barrier and lenticular lens array [Masia et al. 2013]. Although these methods can offer effective 3D images, they require calculation and generation of precise images for multiple viewpoints, and users have to stay within a limited view angle.

A different approach to realize advanced 3D displays is using a physical 3D space to render graphics instead of a planar surface and forming a visual representation of an object in three physical dimensions, as opposed to the planar image of traditional screens that simulate depth through various visual effects [Masia et al. 2013]. These 3D displays, which are called volumetric displays, allow users to view the displayed images from any angle. Volumetric displays arrange "voxels" in a 3D space. They are divided into two categories by the characteristics of the voxels: emitting or reflecting light. The voxels emitting light may be LEDs [Clar 2008], end points of optical fibers [Willis et al. 2012], or laser-induced plasma [Kimura et al. 2006]. Those reflecting projected light may take the form of fog [Rakkolainen et al. 2005a], water drops [Barnum et al. 2010], or floating small particles [Ochiai et al. 2014]. In this study, we focus on laser-induced plasma.

Laser-induced plasma has the following advantages. First, it does not require physical matter arranged and suspended in air to emit light. Second, it does not require wires and structures that possibly obstruct the line-of-sight because power is transmitted wirelessly. Third, the laser can be precisely controlled owing to the progress in optical technologies.

We envision a laser-induced plasma technology in general applications for public use. If laser-induced plasma aerial images were made available, many useful applications such as augmented reality (AR), aerial user interfaces, volumetric images could be produced (Figure 2). This would be a highly effective display for the expression of three-dimensional information. Volumetric expression has considerable merit because the content scale corresponds to the human body; therefore, this technology could be usefully applied to wearable materials and spatial user interactions. Further, laser focusing technology adds an additional dimension to conventional projection technologies, which offer surface mapping while laser focusing technology is capable of volumetric mapping. Thus, this technology can be effectively used in real-world-oriented user interfaces.

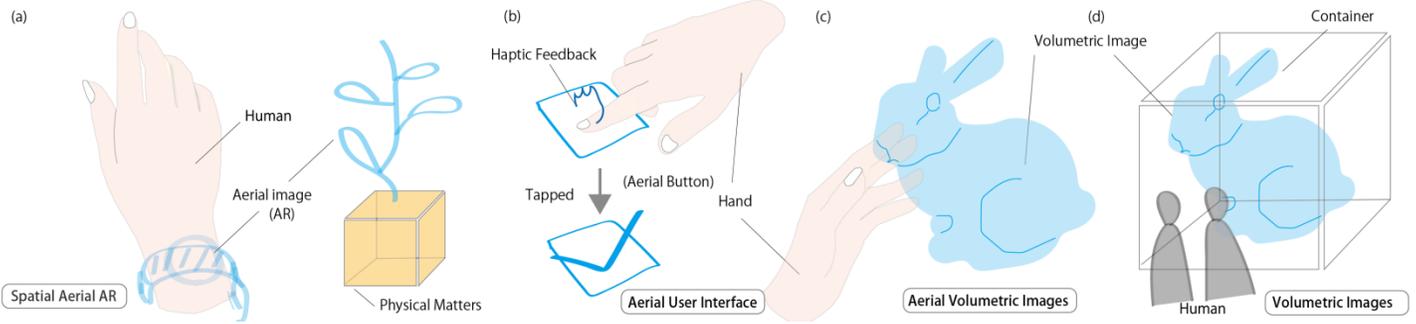

**Figure 2:** *These figures show the example applications of proposed laser-based graphics technology. (a) Images superposed on a hand and a box. (b) Floating button with haptic feedback. (c-d) Volumetric images rendered in open and closed areas.*

Plasma-based 3D displays were previously developed using a nanosecond laser [Kimura et al. 2006] and femtosecond (100fs) laser [Saito et al. 2008]. These studies on laser-plasma graphics were pioneering but still uncompleted. Our motivation is to expand their achievements and provide complete discussion on this laser-plasma graphics technology.

In this study, we use femtosecond lasers with pulse durations of 30-100 fs, and 269 fs. This leads to safer plasma generation than nanosecond lasers, which can be incorporated into our daily lives. The design space and possible scenarios of the plasma-based 3D display are discussed. In addition, we use an optical device, called the spatial light modulator (SLM), to modify the phase of light rays and produce various spatial distributions of light based on interference.

The primary contribution of this paper is the production of an in-air SLM-based laser-plasma graphics that enables physical contact and interaction by ultra-short pulse duration laser. Also, the principles are theoretically described, the characteristics of this technology are experimentally examined, and the applications and scalability are discussed.

The remaining sections of this paper discuss the following: First, we describe the principles and design parameters of femtosecond-laser-based volumetric displays. We explore a safe, high-resolution, wide-variety laser-based volumetric display using a femtosecond laser and an SLM. Second, we introduce the setup we designed. Third, we give examples of applications. Finally, we conduct experiments on generation, safety, and control of lasers. We also discuss the limitations and estimate scalability. We believe that this study fills the gaps in design space of plasma-based 3D displays that were left unresolved by previous studies.

## 2 Related work

| | Non position control | Position control |
|---|---|---|
| **Reflection** | Mechanical motion of mirror or screen [Paker] [Jones] [Favalora] [Karnik] | Screens deformed by linear actuators [Iwata] [Follmer] |
| | Fog [Rakkolainen] [Lee] | |
| | Water drops [Eitoku] [Barnum] | Soap film deformed by focused ultrasound [Ochiai] |
| | Liquid crystal shutters [Sullivan] | |
| | Photochromic materials [Hashida] | Small particles by acoustic levitation [Ochiai] |
| | Floating small particles [Perlin] | |
| | Launched particles [Matoba] | |
| **Emission** | Cubic array of LEDs [Clar] | Moved optical devices [Jansson] |
| | 3D fabricated optical fibers [Willis] [Pereira] | Fluorescence voxels [Macfarlane] |
| | | LEDs on linear actuator [Poupyrev] |
| | | Laser plasma [Kimura] |
| | | <span style="color:red">This study</span> |

**Figure 3:** *A map of related work divided into four categories regarding to non-position-control/position control and reflection/emission. This study falls into the position-control and emission category.*

In this section, first, we survey conventional studies on volumetric displays and divide them into four categories: Non-position-control/position-control and reflection/emission. Note that some of them are not 3D but 2.5D, and position-control in 2.5D means surface deformation. Our study is associated with the position-control and emission category (Figure 3). Next, previous studies on plasma-based 3D display are described, and the issues that have not been discussed are pointed out. Last, studies on aerial interaction are cited, and an additional property of our study is clarified.

### 2.1 Volumetric displays

#### 2.2.1 Non-position-control types

Reflection: In this category, the work space is filled with small objects of a material that can passively reflect projected or environmental light. 3D displays based on mechanical motion of mirror or screen are discussed in [Parker 1948]. A spinning mirror is used with a high-speed projector in [Jones et al. 2007], where different images are projected onto the mirror according to its azimuthal angle to express a 360± light field of an object. Similarly, images are projected onto a rotating screen [Favalora et al. 2002] and a rotating diffuser plate [Karnik et al. 2011]. The systems proposed in [Rakkolainen et al. 2005a; Lee et al. 2009] use fog as reflecting material. A thin layer of fog is generated and images are projected onto it. In [Eitoku et al. 2006], falling water drops are utilized as a screen. The lens-like property of water drops delivers projected images to users' eyes. Subsequently, multilayer water drops screens were implemented [Barnum et al. 2010] and different images were projected onto different layers by synchronizing the projector with the water valves. In DepthCube [Sullivan 2004], a multi-layered liquid crystal shutters is illuminated by a high-speed projector. Photochromic materials are used in [Hashida et al. 2011] to form a volumetric and multi-color display controlled by an ultraviolet projector. Holodust [Perlin and HAN 2006] illuminates floating small particles by lasers. Small particles are launched into air and illuminated by a projector in [Matoba et al. 2012].

Emission: In this category, objects occupying the work space actively emit light to show images. Clar [Clar 2008] created a 3D cubic array of LEDs. In this setup, the LEDs are supported by a framework and the relative positions of them (i.e., voxels) are fixed. Currently fabrication type 3D volume displays are explored [Willis et al. 2012]. 3D print objects with embedded light paths can display information when the objects are placed on a flat display. As the objects get more complex, the light paths also get complex and make it difficult to design the object to be printed. Pereira et al. [Pereira et al. 2014] solve the issue by algorithmically computing the arrangement of the light paths so that their endings form a desired surface shape, such as that of a face.

### 2.1.2 Position-control types

Reflection: In this category, the positions of reflection objects are controlled to render graphics. Studies focusing on controlling the surface shape of a screen or display have also been pursued. For example, the deformable screen Project FEELEX [Iwata et al. 2001] changes its surface shape by linear actuators. A deformable screen inForm [Follmer et al. 2013] not only displays images on it but also interacts with objects. [Ochiai et al. 2013] used focused ultrasound to deform a soap film, without making contact, to show a bump on it. Pixie Dust [Ochiai et al. 2014] is a floating display consisting of small particles that are suspended and moved by means of acoustic levitation.

Emission: Light sources are moved to realize 3D displays in this category. This type of volumetric displays was originally reported in [Jansson and Berlin 1979]. Many types of volumetric displays are explored for 35 years. [Grossman and Balakrishnan 2006] did great survey on this volumetric area. [Macfarlane 1994] proposed a voxel-based spatial display. LUMEN [Poupyrev et al. 2004] comprises of LEDs attached to linear actuators and shows information in the form of RGB (red-green-blue) and H (height). Laser plasma, which is free from physical support and connection, is used as a light source in [Kimura et al. 2006]. We also work in this technology to use this advantage.

### 2.2 Laser-based volumetric displays

As mentioned before, laser-plasma 3D displays are categorized as the position-control and emission type 3D display. Voxels in air are generated by high-intensity lasers which are achieved by shortening pulse duration (e.g. nanoseconds or shorter) under a limited total power.

The basic concept was demonstrated using a nanosecond laser in [Kimura et al. 2006] where a rendering speed was 100 dot/sec. Later, 1,000 dot/sec was achieved [Saito et al. 2008] by adoption of a femtosecond (100 fs) laser[1]. The color of voxels was bluish white because of plasma emission. With the latter one, rendering algorithms of point cloud were discussed in [Ishikawa and Saito 2008a; Ishikawa and Saito 2008b]. Although these studies on laser-plasma graphics were pioneering, the detailed discussion on the light emission, design space, scalability, and so on was not provided in the published papers. We provide the discussion on these issues in this paper, and complete laser-based graphics in air from principles to applications.

Laser-based 3D displays in materials other than air were also demonstrated. An in-water type[2] of laser-based volumetric display was developed in [Kimura et al. 2011] where 50,000 dot/sec was achieved. While no detailed principle was provided, we infer that this in-water type is not based on laser plasma but laser-induced microbubbles. The green light dots generated by a green laser can be explained as diffusion of the incident laser by the microbubbles. Fluorescent materials were used in [Soltan et al. 1992; Downing et al. 1996; Hasegawa and Hayasaki 2013]. Pulse peak intensity required for the microbubble- and fluorescence-based rendering, as experimentally confirmed in Section 5.3. This offers higher rendering speed than the plasma-based rendering so that not a set of lines but a surface can be represented [Ishikawa et al. 2011].

### 2.3 Aerial interaction

Volumetric, aerial, and/or 3D displays are usually accompanied by interaction with users' hand. For example, users can directly interact with graphics rendered on a thin layer of fog [Rakkolainen et al. 2005b]. Touchable Holography [Hoshi et al. 2009] and RePro3D [Yoshida et al. 2010] show 2D and 3D images in air, respectively, and also provide haptic feedback. Small particles are acoustically levitated in [Ochiai et al. 2014] and users can touch them. ZeroN [Lee et al. 2011], although it is a tangible system rather than a graphic system, magnetically levitates a sphere and users can touch and also handle it. For aerial interaction, there are two necessary conditions on volumetric displays. They should be safe and accessible. The previous works based on lasers do not satisfy these conditions. The in-air type [Kimura et al. 2006] is harmful to users' hand because of plenty of energy and in-water type [Kimura et al. 2011] renders images in a transparent container filled with water. In this paper, we attempt to demonstrate safe and accessible laser-based volumetric display.

Table 1: *Comparison between the previous and this study.*

|  | Kimura 2006 | Saito 2008 | This study |
|---|---|---|---|
| Medium | Air | Air | Air |
| Laser | Nanosecond | Femtosecond 100 fs | Femtosecond A: 30-100fs, 2mJ B: 269fs, 50μJ |
| Dot/sec | 100 | 1,000 | A: 1,000 B: 200,000 |
| Hologram | No | No | Yes LCOS-SLM or LCSLM |
| Interaction | No | No | Yes |
| Touch | No | No | Yes |

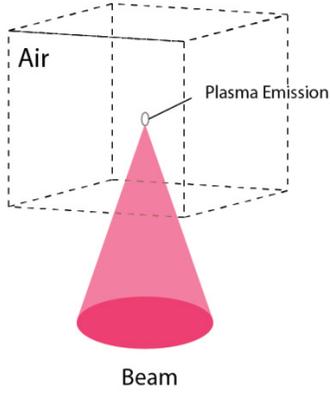 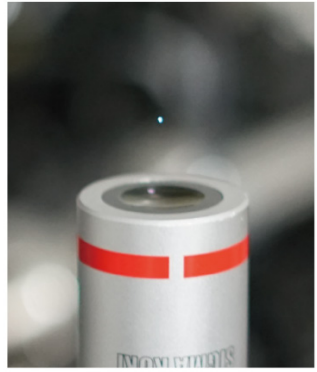

**Figure 4:** *Laser plasma induced by focused femtosecond laser.*

## 2.4 Position of this study

The development of a volumetric display has two problems which have been encountered in conventional studies: how to suspend and emit voxels. The application of laser plasma technology to a volumetric display overcomes these two issues, because laser plasma generates an emission point at an arbitrary position in a 3D space. In addition, studies on conventional laser volumetric displays have not sufficiently discussed theoretical principles and scalability. This study focuses on a system for rendering volumetric graphics in air using a femtosecond laser. An ultrashort-pulse laser and SLM are used in our system, which allows us to explore touch interaction and computer-generated holograms. These explorations and evaluations are useful as regards discussion of the scalability and application space of a plasma-based volumetric display using a high-intensity laser for general, wide-spread application.

## 3 Principles

In this section, we show how to generate light spots by lasers.

### 3.1 Laser-induced light spot

There are three types of laser-induced effects (Figure 5) that produce light spots, and fluorescence is one among them. First, an orbital electron in a molecule or atom is excited when the atom absorbs one or more photons. Next, a new photon is emitted when the electron relaxes. If two photons are absorbed at the same time, the wavelength of the emitted photon is half of that of the original photons. The wavelength required to excite an electron is dependent upon the type of fluorescent material. The emitted light has $N$ times shorter wavelength when $N$ photons are absorbed simultaneously. This effect occurs with a relatively low-intensity laser (an energy of nJ to mJ is sufficient). Confocal laser microscopy is based on this effect [Denk et al. 1990; Paddock 1999].

Cavitation is another effect that plays a key role. Microbubbles are generated at the focal point of a laser in a liquid medium. This localized cluster of microbubbles diffuses the incident laser such that the laser is seen as a point light. The color of this point light depends directly on the wavelength of the incident laser. This fact indicates that RGB images can be expressed by using multiple lasers of different wavelengths. The microbubbles show just environment light if the laser is invisible (infrared and ultraviolet). This effect requires an intense laser to generate microbubbles.

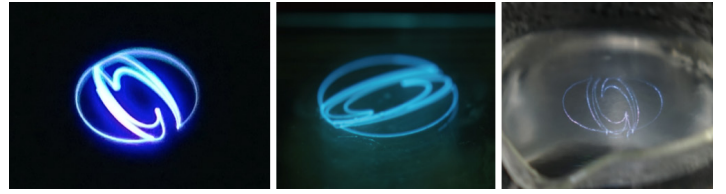

**Figure 5:** *Image examples rendered with a 269-fs infrared laser. (Left) Laser plasma in air by Galvano scanning. (Center) Fluorescent emission in fluorescent solid material by CGH. (Right) Microbubbles in water by Calvano scanning.*

The last effect is plasma, or ionization. In particular, tunnel ionization can produce sufficiently visible light, which dominantly occurs when the laser intensity is greater than $10^{14}$ W/cm² [Keldysh 1965]. The potential well of a molecule or atom is deformed by the electric field of the high-intensity laser to have a potential barrier, and then, an electron has the opportunity to leave the atom (i.e., ionization) based on the tunnel effect. It is known that higher laser intensity leads to higher tunnel-ionization probability; that is, more electrons are ionized [Ammosov et al. 1986]. The ionized electron is recombined with the atom after a half cycle and a photon is emitted. This effect is called laser breakdown. The emitted light looks bluish white.

In this study, we focus on the third effect, i.e. ionization, because it can be achieved in air (Figure 4).

### 3.2 Laser filamentation

An emission dot generated by a high-intensity laser has a tail along the propagation direction. This tail is generated as the self-focusing behavior, due to the optical Kerr effect, competes with the natural diffraction of the laser beam; however, this effect is undesirable when rendering 3D graphics in air. Practically, this effect is invisible to the human eye because the light from the focal point is relatively much brighter, but might be taken into consideration in some special cases.

### 3.3 Voxel sizes

We assume that the size of an emission dot (i.e., a voxel) is equal to the size of the focal point of the laser. The focal point is usually an oval that has two diameters. One is the diameter perpendicular to the laser beam, $w_f$, which is the diffraction limit and determined by the original beam width, $a$, the focal length, $r$, and the wavelength, $\lambda$, such that

$$w_f = 2\lambda \frac{r}{a}, \tag{1}$$

The other is the diameter along the laser beam, $w_d$, which is geometrically obtained from the relationship $a : w_f = r : w_d/2$, such that

$$w_d = 4\lambda \left(\frac{r}{a}\right)^2. \tag{2}$$

### 3.4 Computational phase modulation

The use of SLMs is one method to render holograms. In general, an SLM has an array of computer-controlled pixels that modulate a laser beam's intensities, phases, or both. This optical device is used in, for example, laser processing to generate an arbitrary pattern of laser [Hayasaki et al. 2005].

A liquid crystal SLM (LCSLM) is used in this study, which contains a nematic liquid crystal layer. The molecule directions within this layer are controlled by electrodes, i.e., pixels, and the phase of light ray reflected by each pixel is modulated according to the direction of the liquid crystal molecule. In other words, this device acts as an optical phased array.

The spatial phase control of light enables the control of focusing position along both the lateral (XY) and axial (Z) directions. A complex amplitude (CA) of the reconstruction from the computer-generated hologram (CGH) $U_r$ is given by the Fourier transform of that of a designed CGH pattern $U_h$:

$$U_r(\nu_x, \nu_y) = \iint U_h(x, y)\exp\left[-i2\pi(x\nu_x + y\nu_y)\right]dxdy \quad (3)$$
$$= a_r(\nu_x, \nu_y)\exp\left[i\varphi_r(\nu_x, \nu_y)\right]$$

$$U_h(x, y) = a_h(x, y)\exp\left[i\varphi_h(x, y)\right] \quad (4)$$

where $a_h$ and $\varphi_h$ are the amplitude and phase of the hologram plane displayed on the SLM, respectively. In the experiment, $a_h$ is constant because an irradiation light to the CGH is considered as the plane wave with a uniform intensity distribution. $\varphi_h$ is designed by ORA algorithm. On the other hand, $a_r$ and $\varphi_r$ are the amplitude and phase of the reconstruction plane, respectively. The spatial intensity distribution of reconstruction is actually observed as $|U_r|^2 = a_r^2$.

In the control of focusing position along the lateral (XY) direction, the CGH is designed based on a superposition of CAs of blazed gratings with variety of azimuth angles. If the reconstruction has $N$-multiple focusing spots, CGH includes $N$-blazed gratings. In the control of focusing position along the axial (Z) direction, a phase Fresnel lens pattern

$$\varphi_p(x, y) = k\frac{x^2 + y^2}{2f}$$

with a focal length $f$ is simply added to $\varphi_h$, where $k = 2\pi/\lambda$ is a wave number. In this case, the spatial resolution of the SLM determines the minimum focal length, following the theory discussed in Section 3.3.

ORA method is an optimization algorithm to obtain the reconstruction of CGH composed of spot array with a uniform intensity (Figure 6). It is based on adding an adequate phase variation calculated by an iterative optimization process into the CGH. In the $i$-th iterative process, amplitude $a_h$ and phase $\varphi_h^{(i)}$ at a pixel $h$ on the CGH plane, and a complex amplitude (CA) $U_r^{(i)}$ at a pixel $r$ corresponding to focusing position on the reconstruction plane are described in the computer as follows,

$$U_r^{(i)} = \omega_r^{(i)}\sum_h u_{hr}^{(i)} \quad (5)$$
$$= \omega_r^{(i)}\sum_h a_h \exp\left[i\left(\varphi_{hr} + \varphi_h^{(i)}\right)\right]$$

where $u_{hr}$ is CA contributed from a pixel $h$ on the CGH plane to a pixel $r$ on the reconstruction plane, $\varphi_{hr}$ is a phase contributed by the light propagation from a pixel $h$ to a pixel $r$, $\omega_r^{(i)}$ is a weight coefficient to control the light intensity at pixel $r$. In order to maximize a sum of the light intensity $\Sigma r \, |U_r^{(i)}|^2$ at each pixel $r$, the phase variation $\Delta\varphi_h^{(i)}$ added to $\varphi_h^{(i)}$ at pixel $h$ is calculated using flowing equations.

$$\Delta\varphi_h^{(i)} = \tan^{-1}\left(\frac{S_2}{S_1}\right), \quad (6)$$

$$S_1 = \sum_r \omega_r^{(i)} a_h \cos\left(\varphi_r - \varphi_{hr} - \varphi_h^{(i)}\right), \quad (7)$$

$$S_2 = \sum_r \omega_r^{(i)} a_h \sin\left(\varphi_r - \varphi_{hr} - \varphi_h^{(i)}\right), \quad (8)$$

where $\omega_r$ is the phase at pixel $r$ on the reconstruction plane. The phase of CGH $\varphi_h^{(i)}$ is updated by calculated $\Delta\varphi_h^{(i)}$ as follows.

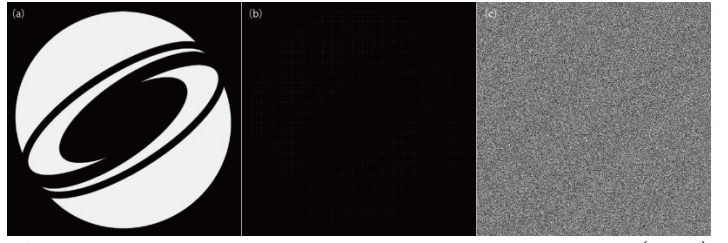

**Figure 6:** *Example of a computer-generated hologram (CGH). (a) An original image, (b) a converted spot-array image of the original image, and (c) a CGH to be displayed on the SLM.*

$$\varphi_h^{(i)} = \varphi_h^{(i-1)} + \Delta\varphi_h^{(i)}, \quad (9)$$

Furthermore, $\omega_r^{(i)}$ is also updated according to the light intensity of the reconstruction obtained by the Fourier transform of Eq. (9) in order to control the light intensity at pixel $r$ on the reconstruction plane.

$$\omega_r^{(i)} = \omega_r^{(i)}\left(\frac{I_r^{(d)}}{I_r^{(i)}}\right)^\alpha \quad (10)$$

where $I_r^{(i)} = |U_r^{(i)}|^2$ is the light intensity at pixel $r$ on the reconstruction plane in the $i$-th iterative process, $I_r^{(d)}$ is an desired light intensity, and $\alpha$ is constant. The phase variation $\Delta\varphi_h^{(i)}$ is optimized by the above iterative process (Eqs. (6)-(10)) until $I_r^{(i)}$ is nearly equal to $I_r^{(d)}$. Consequently, ORA method allows us to design the CGH with the high quality.

### 3.5 Graphics positioning

The galvano mirror used in this study covers an area of $10 \times 10$ mm². Besides, the SLM also renders graphics within the approximately same area. This means that we have two options to place a point at an intended position: One is leading a laser there by the galvano mirrors and the other is modifying the spatial distribution of the laser by the SLM. The conditions and/or response times of these devices determine which is suitable.

The theoretical rendering limit is 33 dots/s for 30 frame/s, because the femtosecond laser is pulsed at a frequency of 1 kHz. The SLM is used to render additional dots in a single frame, while the galvano mirror is used primarily for positioning the rendered holograms.

### 3.6 Spatiotemporal resolution

The number of dots per frame (*dpf*) is a parameter that must be evaluated for laser-based volumetric displays. We now assume the dots are displayed in darkness; therefore, the minimum required energy for each dot is equal to the laser breakdown threshold, $E_{lbd}$. The total output energy, $E_{tot}$, is divided among the dots by the SLM. The number of dots per laser pulse, $N_{dot}$, is expressed as

$$N_{dot} = \frac{E_{tot}}{E_{lbd}}. \quad (11)$$

The number of dots per frame is determined by $N_{dot}$, the repeat frequency, $F_{rep}$, of the laser pulses, and the frame time, $T_f$, which is determined based on the persistence of human vision. Hence,

$$dpf = N_{dot} \times F_{rep} \times T_f. \quad (12)$$

For example, if $N_{dot} = 100$, $F_{rep} = 1$ kHz, and $T_f = 100$ ms, an animation of 10,000 dpf is played in 10 fps. Note that, in practice, the number of dots per frame is determined by the bottleneck of the time response of the galvano mirrors and/or the SLM, instead of by $F_{rep}$.

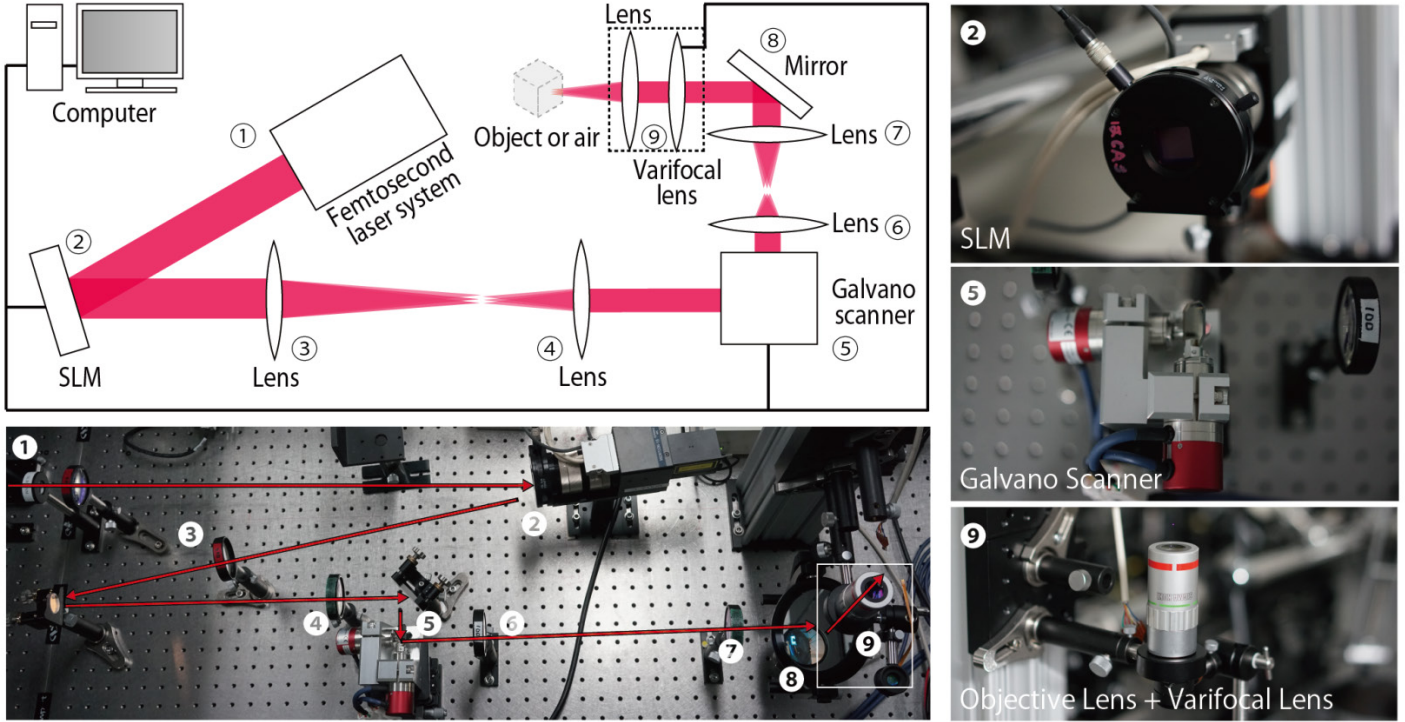

**Figure 7:** *Setup of our light circuit. The host computer controls (2) the SLM for hologram generation, (5) the galvano scanner for XY control, and (9) the varifocal lens for Z control.*

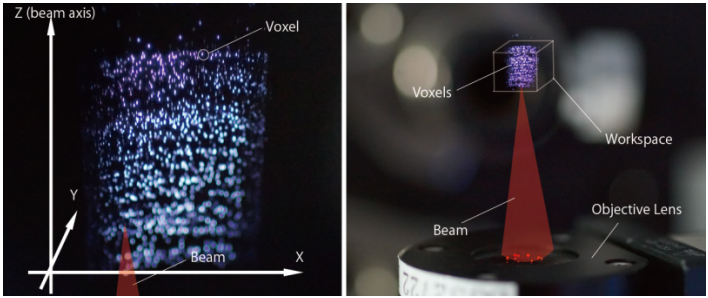

**Figure 8:** *Relationship between the XYZ-coordinate and the focused laser beam. Voxels are rendered above the objective lens.*

## 4 Implementation

In this section, we show our system implementation. First, we introduce an overview of our system. Next, we describe our light source, optical circuit (i.e., arrangement of optical devices), 3D scanning system, SLM, and the control system.

### 4.1 Overview

Figure 7 shows the system configuration of our basic setup. This system aims to produce a simultaneous-multi-point volumetric display. It consists of a femtosecond laser source, an XYZ scanner (galvano scanner + varifocal lens), and a liquid crystal on silicon SLM (LCOS-SLM) displaying a CGH for simultaneously addressed voxels. Our system was tested and investigated at 20.5 deg C. The atmosphere was ordinary air (80% $N_2$ and 20% $O_2$).

The setup was tested using three light sources (A and B), the specifications of which are given below. We primarily used a femtosecond laser source developed by Coherent Co., Ltd., which has a center wavelength of 800 nm, a repetition frequency of 1 kHz, and pulse energy in the 1 to 2-mJ range. The specifications of the laser sources are shown in Table 2. Figure 8 shows example results for our system in the air.

The galvano mirror scans the emission dot along the lateral directions (X- and Y-scanning), while the varifocal lens can vary its focal point in the axial direction (Z-scanning). The Fourier CGH is used for simultaneously addressed voxels [Hayasaki et al. 2005]. The CGH, designed with an optimal-rotation-angle (ORA) method [Bengtsson 1994], is displayed on the LCOS-SLM, which has 768 × 768 pixels, a pixel size of 20 × 20 $\mu m^2$, and a response time of 100 ms. The specifications of each component are shown in Tables 3 and 4. In addition to these components, we use a microscope for monitoring and recording which is connected to the computer via USB.

### 4.2 Light source

We use two light sources. The light source that is primarily used for evaluation and application was developed by Coherent Co., Ltd and has a center wavelength of 800 nm, repetition frequency of 1 kHz, pulse energy of up to 2 mJ, and the pulse width is adjustable from 30 to 100 fs. Figure 9 shows the spectra and pulse intensities of the 30- and 100-fs settings with this light source. Ultra-short pulses are generated by converting low-intensity and long-duration pulses to high-intensity and short-duration ones. If the average laser pulse energy is unchanged, the peak intensity differs according to pulse width. In fact, the 30-fs pulse width has a three-fold greater peak intensity than the 100-fs pulse width at the same average energy. We refer to the system using this light source as System A.

**Table 2:** *Specifications of laser sources.*

| System | A | B |
|---|---|---|
| Maker | Coherent | IMRA |
| Pulse duration | 30-100 fs | 269 fs |
| Repeat cycle | 1 kHz | 50 kHz |
| Energy/pulse | 2 mJ | 50 $\mu$J |
| Dots/sec | 1,000 | 200,000 |
| Average power | 2 W | 10 W |

The other light source we used is the FCPA µJewel DE1050 from IMRA America, Inc. The laser has a center wavelength of 1045 nm, repetition frequency of 200 kHz, pulse energy of up to 50 µJ and pulse width of 269 fs. We refer to the system with this light source as System B. Note that the peak intensity of laser is important to produce the aerial plasma, rather than the pulse width. Both of Systems A and B have sufficient peak intensity to excite the air and generate emission dot.

### 4.3 Optical circuit

Here, we describe our optical circuit following the path of the laser. Figure 7 shows the optical setup of System A. The laser is generated by the femtosecond light source and then phase-modulated by the SLM. The SLM energy conversion rate is 65 to 95%. Then, the beam spot is varied by two lenses (F = 450 and 150 mm). Through this two-lens unit, the beam spot is reduced by a factor of 1/3. It is then reflected by the galvano mirror, which determines the XY-position of the light. The galvano and SLM are connected in an object-image correspondence. Subsequently, the beam spot is adjusted by two lenses (F = 100, 150 mm); this two-lens unit magnifies the beam spot 1.5-fold. Then, the light enters the varifocal lens. The varifocal lens and galvano mirror are connected in an object-image correspondence and the former adjusts the z-axis focal points. The light enters the objective lens (F = 40 mm). Once it exits this lens, it excites the display medium (air). The energy conversion rate of System A is 53%.

System B has the same structure but lacks a SLM. Also, the lens sets are slightly different from those of System A. Specifically, System B has no lens before the galvano mirror, as the varifocal lens is positioned after the galvano mirror. Then, the beam spot is adjusted by the two-lens unit (F = 50, 80 mm). An F20 objective lens is employed and System B's total energy conversion rate is 80%.

### 4.4 3D scanning system

In this subsection, we describe our scanning system in detail. Figure 7 shows the galvano and varifocal lenses. We employ galvano mirrors to scan the lateral directions (X- and Y-scanning), while a varifocal lens can change its focal point in the beam axial direction (Z-scanning). For system A, we utilize a Canon GH-315 driven by GB-501 as the galvano mirror and for System B we employ an Intelliscan 20i to scan the beams. Both are connected by PCI boards. Table 3 shows the specifications of each of the galvano mirrors. We employ an Optotune EL-10-30 for both Systems A and B as the varifocal lens, which is connected via USB serial to a PC. The specifications of the varifocal lens are shown in Table 4. These devices are operated by original applications coded in C++.

### 4.5 LCSLM

The LCSLM (Hamamatsu, PPM) is a parallel-aligned nematic liquid crystal spatial light modulator (PAL-SLM) coupled with a liquid crystal display (LCD) and a 680-nm laser diode (LD). This device, which can perform phase-only modulation of more than 2 radian, is frequently used to display real-time CGHs. The PALSLM is composed of a liquid crystal (LC) layer, a dielectric mirror, and an optically addressed photoconductive (PC) layer containing amorphous silicon, which are sandwiched between two transparent indium tin oxide electrodes. The LC molecules are aligned in parallel. When incident light illuminates the PC layer, the impedance of this layer decreases and the electric field across the LC layer increases accordingly. With this increased field, the LC molecules become tilted in the propagation direction of the readout light and the effective refractive index of the LC layer decreases. Pure phase modulation occurs only when the polarization direction of the femtosecond laser is parallel to the aligned direction of the LC molecules. The CGH pattern on the LCD illuminated by the LD is applied to the PC layer through an imaging optics.

### 4.6 Control system

Figure 7 shows our system diagram. The system is controlled using a Windows PC operating system, with all programs coded in C++. The control system operates the SLM, galvano mirror, and varifocal lenses. To monitor the interaction, a USB microscope is connected to the system. The galvano and varifocal lenses run along different threads and are synchronized when new draw patterns are input. The user input is captured at 20 Hz and the SLM is connected to the computer as an external display.

## 5 Experiments and Evaluations

In this section, we describe our experiments and system evaluation procedures. Firstly, we introduce an overview of our experimental plan and results. Then, we report the results of the following tests: Energy vs ionized plasma brightness, brightness vs pulse peak, simultaneously addressed voxels for aerial images, and skin damage. In the experiments, the brightness are measured as a summation of all the pixel values within an close-up image of the plasma taken by a digital camera, which is a common definition in the field of laser optics.

We tested not only gas-ionized plasma, but also photon absorption and cavitation, in order to compare the various energy consumption performances and the means of applying the femtosecond laser system to the display technology. All experiments were conducted using System A, which is described in Section 4.

**Table 3:** *Specifications of galvano mirrors.*

| System | A | B |
|---|---|---|
| Maker | Canon | Intelliscan |
| Scan angle | ±0.17 rad | ±0.35 rad |
| Error | < 5 $\mu$rad | < 5 mrad |
| Resolution | 20 bit | 20 bit |

**Table 4:** *Specifications of varifocal lens.*

| System | A and B |
|---|---|
| Maker | Optotune |
| Aperture | 10 mm |
| Response time | < 2.5 ms |
| Focal length | +45 to +120 mm |

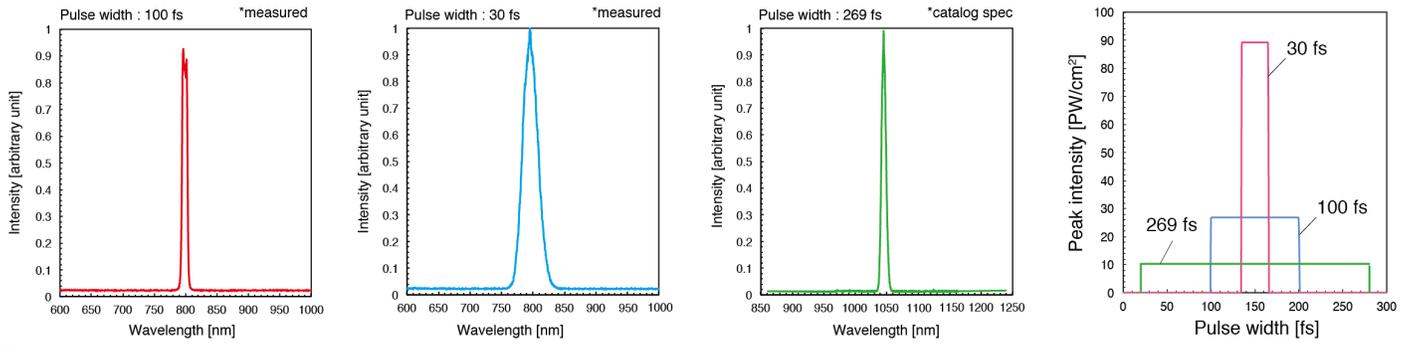

**Figure 9:** *Spectra of 100-fs, 30-fs, and 269-fs lasers (from left to right). The rightmost is the peak intensity and the pulse width of each femtosecond lasers.*

## 5.1 Experiments overview

In this study, we aim to propose a femtosecond laser-based display system design. In conventional studies, the requirements, scalability, and safety of such laser-based systems are not thoroughly discussed. There are several factors that we should explore. In Section 5.2, we examine the voxel brightness, which is important in relation to the energy and display spatiotemporal resolution, as discussed in Section 3. In Section 5.3, we explore the relationship between pulse duration and brightness. This is important for scalability, particularly when a faster laser source is developed. In Section 5.4, we explore simultaneously addressed voxels with SLM. This is important for scalability in accordance with increasing spatiotemporal resolution. Then, in Section 5.5, we examine safety issues and the effect of the plasma on skin. This is important as this technology is intended for widespread, general use. Finally, in Section 5.6, we examine audible sound from the plasma generated by femtosecond lasers. This is important as this technology is intended to be used in our daily lives.

## 5.2 Energy vs Brightness

We conducted this experiment to evaluate the relationship between the plasma-production energy level and the resultant brightness of the image. In conventional studies, the minimum peak intensity necessary to produce the ionized plasma is estimated. However, this experiment aimed to confirm the feasibility of our system and to investigate how it can be applied to display voxels and, thus the minimum peak intensity value was determined.

We conducted the experiments using System A (30 fs) and employed a microscope to capture the resultant image. With our setup, the laser source can provide power of up to 7W, however, unwanted breakdown occurs in the light path before the objective lens under too high power. Hence, the full power of the laser source cannot be used. Moreover, the energy capacity of our SLM is not guaranteed over 2 W. The experiments were conducted for a power range of 0.05 to 1.00 W.

Figure 10 (30 fs) shows the experimental setup and results. The experiments were conducted under energies per pulse of 0.16 to 0.55 mJ. The 30-fs laser can produce plasma from 0.2-mJ pulse energy. The cross-sectional area of the focal point is theoretically calculated to be $2 \times 10^{-7}$ cm². Then, the peak intensity is 36 PW/cm² and surely higher than the ionized plasma threshold (> 1 PW/cm²).

## 5.3 Brightness vs pulse peak

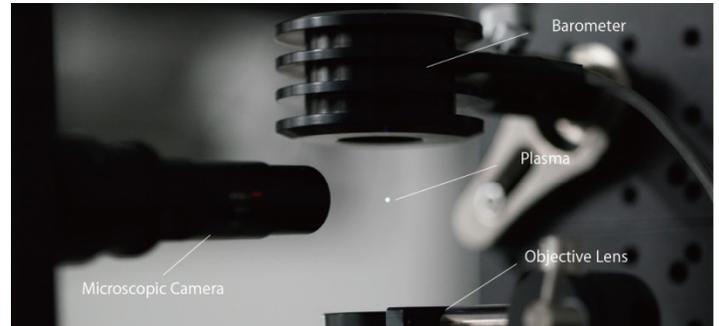

(a) Setup.

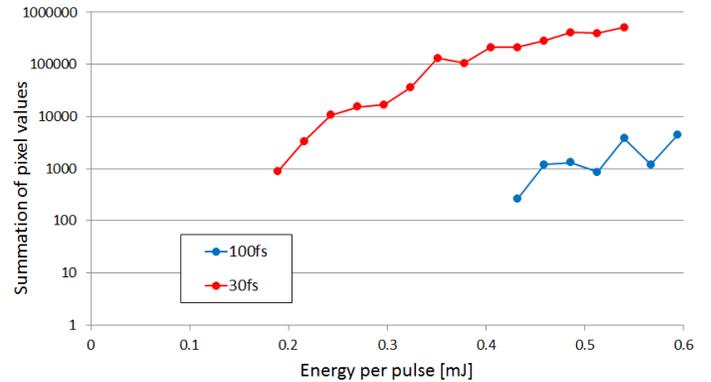

(b) Results.

**Figure 10:** *Experimental setup and results on brightness of light emission in air induced by 30-fs and 100-fs lasers.*

The relationship between the pulse peak and the resultant image brightness was also examined, as the peak intensity plays an important role in plasma generation. This experiment aimed to classify systems of different pulse width in terms of display voxel brightness.

As previously, we conducted experiments using System A (30 and 100 fs). Pulses of 30 and 100 fs yield different spectra and peak energies for the same average powers. Also, the 30-fs setting yields a three-fold higher peak pulse. We employed the same microscope to capture the image that was used in Section 5.1 and the results are shown in Figure 10. The experiments were conducted for a power range of 0.05 to 1.00 W.

As a result, it was found that a 100-fs laser can generate plasma from 0.45-mJ pulse energy. Then, the peak intensity is 24 PW/cm² and surely higher than the ionized plasma threshold (> 1 PW/cm²). Besides, it is confirmed that the 30-fs pulse requires less energy than the 100-fs pulse to produce plasma under the same average power.

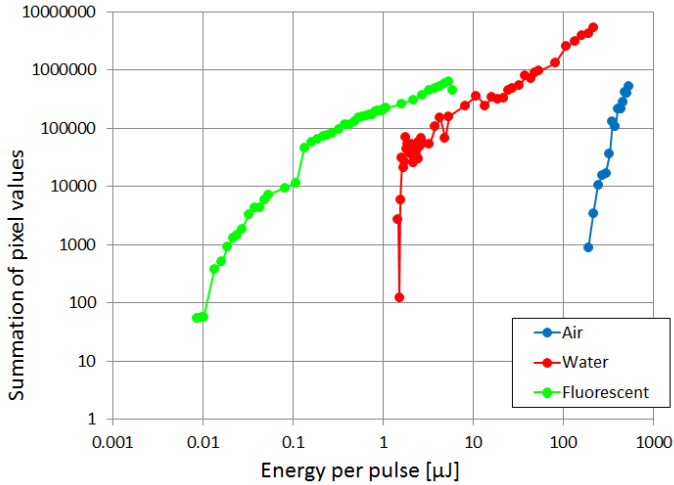

**Figure 11:** *Experimental results on brightness of light emission in air, water, and fluorescence solution induced by 30-fs laser.*

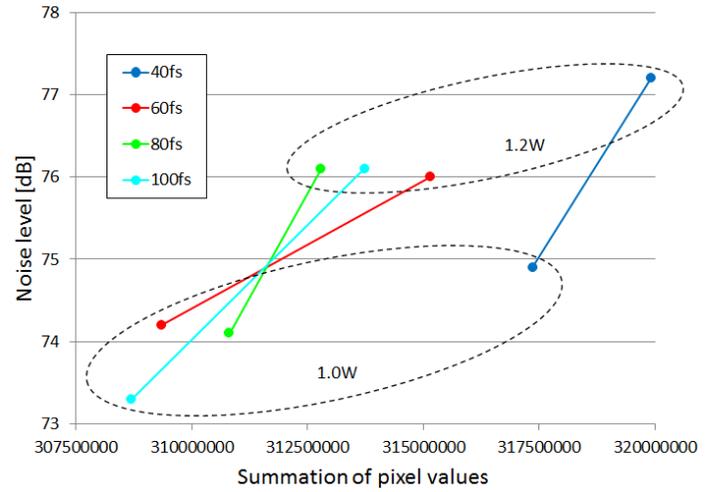

**Figure 14:** *Experimental results on noise level vs. brightness of light emission. The background noise level was 55.7 dB SPL.*

Additionally, we conducted other experiments comparing media materials (air, water, fluorescence solution). The results are shown in Figure 11. It shows that the values of required pulse energy are dramatically different depending on the media materials.

### 5.4 Simultaneously addressed voxels

One of the main contributions of this paper is the application of SLM to in-air laser plasma graphics. This enables simultaneously addressed voxels using CGHs. (Note that, in conventional systems [Kimura et al. 2006; Saito et al. 2008], multiple voxels were not generated simultaneously.) Simultaneous addressing is important to increase the spatiotemporal resolution although the simultaneously addressed voxels are darker than a single point because the energy is distributed among them. This experiment was designed to explore the resolution scalability by using SLM with a single light source. Simultaneous addressing is available for both the lateral (X, Y) and beam (Z) axes, by displaying appropriate holograms on a single SLM. Here, we investigated simultaneous addressing for the lateral axis. Again, the experiments were conducted using System A (30 fs), and Figure 12 shows the results and the holographic images used in the SLM. We employed the same microscope shown in Figure 10. We conducted experiments with a laser power from 0.05 to 1.84 W. We had 1 to 4 simultaneously addressed voxels and 5 or more voxels were not visible.

### 5.5 Skin damage

Another main contribution of this paper is estimating the safety of femtosecond laser systems. Plasma has high energy and can be harmful to humans. However a femtosecond pulse is an ultrashort pulse laser, which is used for non-heat breaking for industrial purposes. It is also used for ultra-short scale fabrication of sub-micrometer order. Thus, we supposed that such pulses may not damage human skin seriously. In addition, our display scans a 3D space very rapidly, therefore, the laser spot does not remain at a specific point for a long period. On the other hand, this plasma still poses dangers for the retina. However, we believe that the potential for general application still exists with appropriate installation.

Therefore, we conducted this particular experiment to explore the damage to skin structure caused by femtosecond plasma exposure. We employed leather for these experiments, as a substitute for human skin.

The experiments were conducted using System A (30 fs and 1 W, 100 fs and 1 W) and the plasma exposure duration was varied between 50 and 6,000 ms. Figure 13 shows the results. It was found that the 30- and 100-fs pulses have almost the same effect on the skin. As we described previously, the 30-fs pulse has a three-fold greater peak energy and can generate brighter voxels. However 50 ms includes 50 shots and there is almost no difference between the 30-fs and 100-fs results. In this experiment, the average power is the factor determining the result. For exposure of under 2,000 ms (2,000 shots), only 100-μm-diameter holes appeared and there was no heat damage to the leather. For a period of longer than 2,000 ms, heat effects were observed around the holes.

We conducted a test with a nanosecond laser for comparison with this result. With the nanosecond laser, the leather burned within 100 ms. This means that pulse duration, repetition times, and energy are important factors affecting the level of damage caused by the laser. Hence, this laser is somewhat safe for use. Further, there are two ways in which the laser can be used safely. One is as an ultra-short-pulse laser, which is bright and has an average output that is not highly intensive. The other is by increasing the scanning speed.

### 5.6 Noise level

The laser plasma in air radiates not only visible light but also audible sound. We conducted an experiment to evaluate the radiated sound. The position of the laser plasma was fixed. The laser power was set at 1.0 and 1.2W. The pulse width was set at 40, 60, 80, 100 fs. The noise level was measured by a noise meter (NL-52, Rion Co., Ltd.), which was placed at 20 mm from the laser plasma. The background noise level was 55.7 dB SPL. The brightness of laser plasma was also recorded.

The results are shown in Figure 14. The maximum noise level was 77.2 dB SPL with 40-fs pulses which also radiate the brightest. This noise level was not so annoying subjectively and it is acceptable in our daily lives. The brighter plasma emission tends to be accompanied by the louder sound. 40-fs pulses radiate louder sound and brighter light.

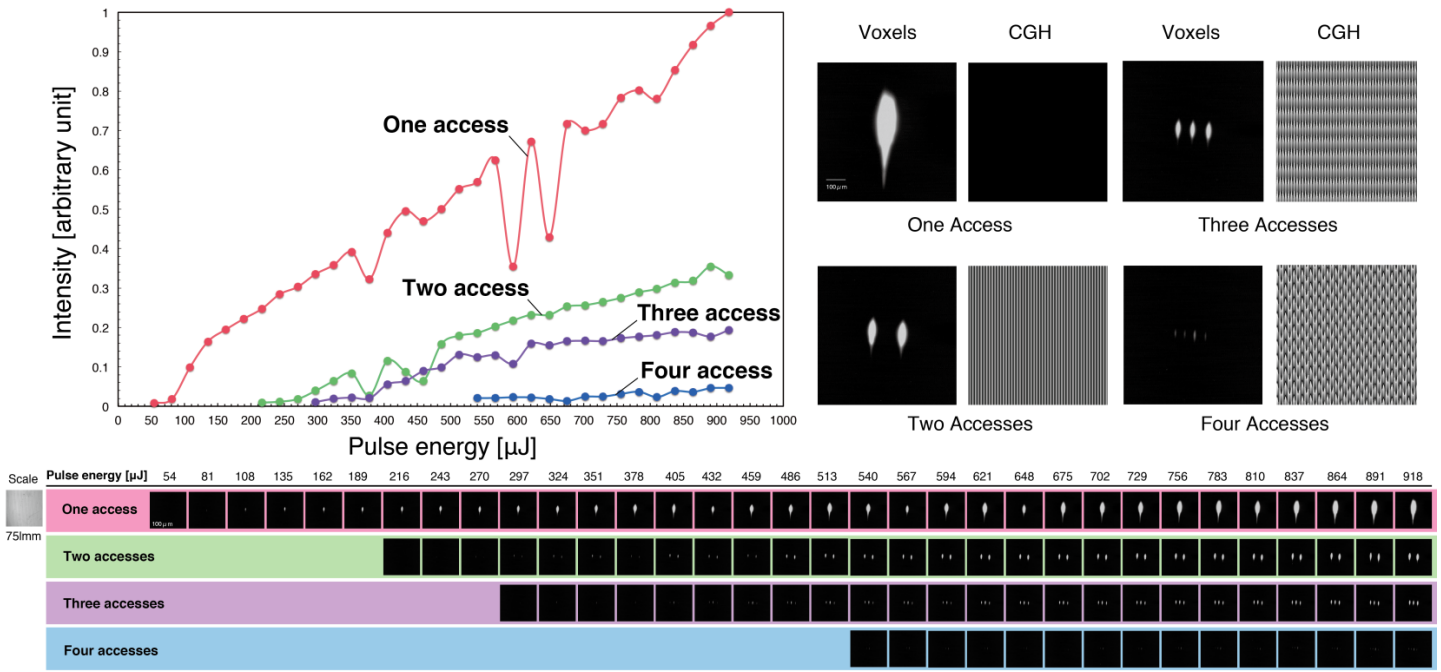

**Figure 12:** *Experimental results on simultaneous addressing. One to four addressing were tested. The intensity is the normalized value of the summation of all the pixel values of the photos of the voxels, which is taken a evaluative value of brightness.*

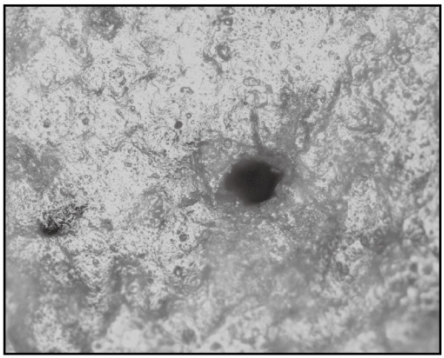

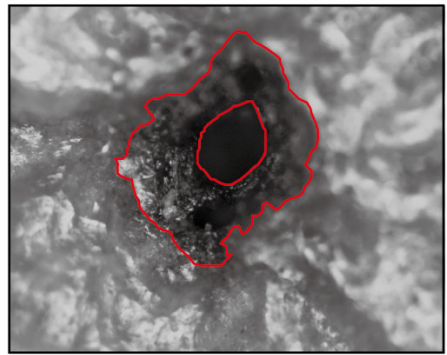

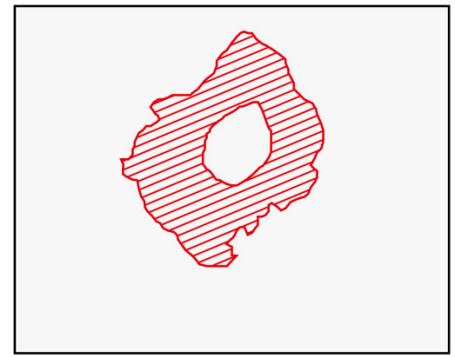

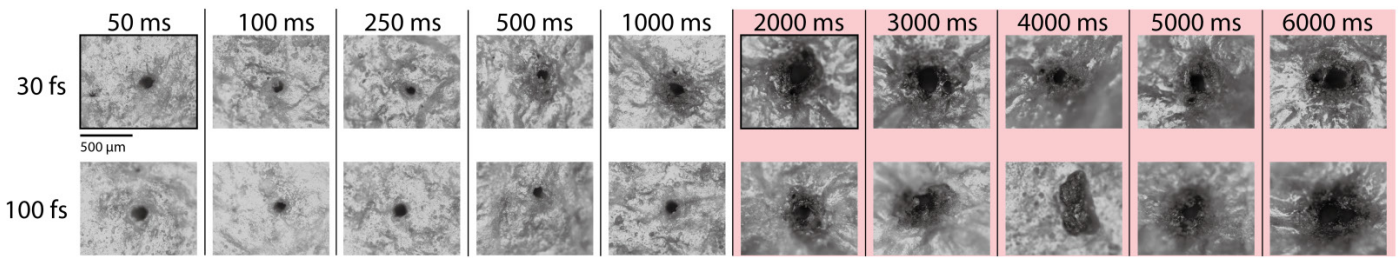

**Figure 13:** *Experimental results on skin damage. Leather sheets were exposed to the 30-fs and 100-fs lasers and the irradiation time was controlled. The exposure longer than 2,000 ms burns the leather surface.*

# 6 Applications

In this section, we describe potential applications of our system. We introduce a 3D aerial display system and interaction between the system and users (Figure 2).

## 6.1 Aerial displays

In this subsection, we describe our aerial display using laser plasma. We developed our application for both Systems A and B and the results are shown in Figures 15 (a), (b), and (d). For Systems A and B, the workspaces are 1 and 8 mm³, respectively. These workspaces are smaller than those of conventional studies, but their resolutions are 10 to 200 times higher than conventional methods. The maximum spatiotemporal resolution is 4,000 point/s (with 4 simultaneous addressing) for

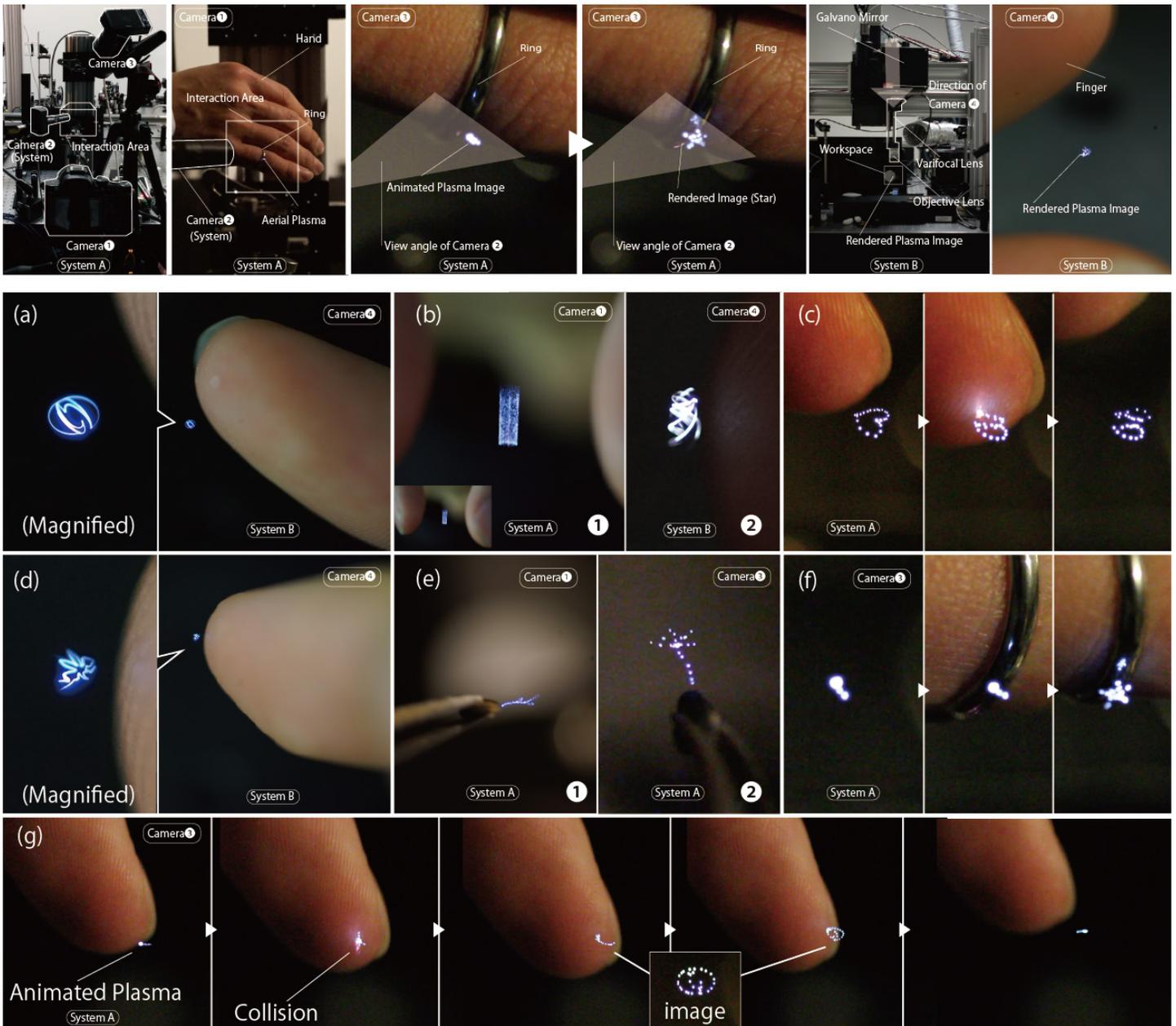

**Figure 15:** *Results of aerial rendering. (Uppermost) The setups of systems A and B. (a) The SIGGRAPH logo, (b) a cylinder, (c) a "heart" that is broken by touch, (d) a "fairy," (e) "sprouts" coming out from seeds, (f) a light point that changes into a "jewel" in contact with a ring, (g) direct interaction between a light point and a finger.*

System A and 200,000 point/s for System B. The image frame rate is determined by the number of vertices used in the image.

### 6.1.1 Spatial AR to real-world object

This aerial display can be used with real-world objects, as shown in Figures 15 (e) and f. One of the merits of the spatial AR to real-world object technique is the AR content is on the same scale as that of the object that is overlapped. Also, this system was developed with a microscope, which can detect an object in the workspace, overlap it with contents, and modify the contents when a contact between the object and plasma occurs. This has an advantage over conventional AR approaches in terms of correspondence to the 3D spatial position. Digital content and information are directly provided in a 3D space instead of a 2D computer display.

### 6.1.2 Aerial interaction with aerial content

Our system has the unique characteristic that the plasma is touchable. It was found that the contact between plasma and a finger causes a brighter light. This effect can be used as a cue of the contact. Figures 15 (c) and (g) show examples of this interaction. One possible control is touch interaction in which floating images change when touched by a user. The other is damage reduction. For safety, the plasma voxels are shut off within a single frame (17 ms = 1/60 s) when users touch the voxels. This is sufficiently less than the harmful exposure time (2,000 ms) determined in section 5.4.

### 6.1.3 Haptic Feedback on Aerial Images

Shock waves are generated by plasma when a user touches the plasma voxels. Then the user feels an impulse on the finger as if the light has physical substance. The detailed investigation of the characteristics of this plasma-generated haptic sensation with sophisticated spatiotemporal control is beyond the scope of this paper.

However, example applications such as "an aerial check box" are at least expected. Figure 15 shows the interaction between the user and the aerial image.

# 7 Discussion

## 7.1 Laser-induced emission phenomena

There are two laser-induced emission phenomena other than plasma emission: Fluorescence and diffusion by cavitation, as introduced in Section 3.1. Both can be applied to displays using the laser-and-SLM system. In this section, the differences between the emission phenomena are explained.

The display medium is the key factor determining the potential interactions. While plasma is generated in air, fluorescence requires fluorescent materials (ink, pigment, etc.) and cavitation requires a fluid medium. The medium also determines the energy that is required to emit light. The order of the required energy decreases from air ($PW/cm^2$), water, to fluorescent materials ($MW/cm^2$).

The available wavelengths also differ in these cases. The plasma color is wavelength-independent and, hence, it is reasonable to use invisible wavelengths, e.g., infrared or ultraviolet. In the case of fluorescence, multi-electron fluorescence is reasonable, in which multiple photons are absorbed by molecules and a single photon with shorter wavelength. Full-color rendering is possible by using multiple fluorescent materials. This is acceptable because the invisible ultraviolet source leaves only the emission visible. On the other hand, when applying cavitation in water, a visible wavelength should be used, because the incoming wavelength is diffused by the microbubbles and observed. This feature leads to full-color rendering with multiple lasers of different colors.

The softness of the medium determines the possible forms of interaction. With aerial plasma in air, a user can insert their hand in to the workspace and touch the plasma. This is also possible with non-fluorescent/fluorescent liquid media. However, in the case of a fluorescent solid medium, the voxels cannot be touched directly.

## 7.2 Drawbacks and limitations

There are some disadvantages and limitations to our systems. As previously explained, SLM is not resistant to an intense laser and, therefore, we cannot use full-range laser power to exhibit images in air. Currently, the SLM technology is popular because of the recent development. We are optimistic that a new kind of SLM system with higher reflectance efficiency will solve this limitation in future. Then more numbers of simultaneously addressed voxels can be generated.

In addition, the optical circuit should be developed and treated carefully. Because our system utilizes high-intensity lasers, ionization may occur on the route of the optical circuit. This also limits the available laser power and damages optical components if ionization occurs. Further, plasma generation is a non-linear phenomenon. These should be well considered for safety.

Also, the focusing and aberration are limitations of our systems. We have to focus the light to make focal points to generate aerial plasma. Then the aperture of the objective lens determines the maximum workspace, which limits the angle range of the galvano mirror. In addition, high-speed variation of the varifocal lens would cause an aberration problem. The characteristics of these lenses are important in developing the optical circuit.

## 7.3 Scalability

### 7.3.1 Size of workspace

Scalability in size of workspace is main concern of our project. Aerial plasma is mainly limited by the objective lens after the varifocal lens. Laser plasma generation needs a laser power of $PW/cm^2$, and the objective lens is required for this purpose. The larger aperture of objective lens permits the larger angle range of the galvano mirror, i.e. XY scanning.

Compare to laser plasma, the laser power to excite fluorescents and water is small and the objective lens is not required. Then, their workspaces are limited by the angle range of galvano mirror and depth range of varifocal lens.

### 7.3.2 Number of voxels

We have to develop three factors to scale up our system for daily applications; increasing power of laser source, shortening pulse width to increase peak power, and increasing scanning speed. These enable us to have an amount of simultaneously addressed and scanned voxels within a single frame, keeping visible and touchable features.

The more laser power leads to more simultaneously addressed voxels. The laser power is limited by the safety on skin, unwanted ionization on the route of optical circuit, and the reflection/transmission characteristics of optical devices.

Shortening pulse width has two benefits. One is a higher repetition frequency (i.e. dots per second), keeping a high peak power required for plasma generation. The other is more safety on human skin because of lower amount of pulse energy with a fixed peak power.

Galvano mirror and varifocal lens have a small room to improve scanning speed. Employing multiple laser systems is one of the solutions to generate multiple voxels.

### 7.3.3 Refresh rate

The refresh rate of this system is determined by the number of simultaneously addressed voxels by SLM, the refresh rate of SLM, the scanning speed of galvano mirror, and the response time of varifocal lens. The galvano mirror is the fastest, more than 1 kHz, and the others work at less than 100 Hz. It is hence reasonable to use galvano mirror mainly. In addition, SLM can multiply voxels if its low refresh rate is acceptable. Then the multiplied voxels move together by galvano scanning.

## 7.4 Safety

Class 4 laser sources are used in this paper. The proposed display system was carefully designed and operated based on the International Electrotechnical Commission (IEC) 60825-1:2014. There are two concerns regarding the safety of laser: the damages on eyes and skin. It should be avoided for users to see the laser beam directly. While laser plasma emits visible light in all directions at the focal point and users can see it safely from the side of the laser beam, it is recommended for users to wear glasses with infrared filters until this display technology is well matured and the safety for eyes is confirmed.

There are a few reports on damage of skin by femtosecond lasers. The minimum visible lesion thresholds for porcine skin for pulsed lasers were evaluated in [Cain et al. 2007]. The ED50 for a femtosecond laser (44 fs, 810 nm, and 12 mm spot size) was determined as 21 mJ from the observation that the lesions by the lasers less than that energy value disappeared at 24 hours after the exposure. The energy (2 mJ and 50 μJ for Lasers A and B, respectively) and spot size (less than 10 μm) are much smaller, and we expect that damages by these femtosecond lasers are negligible. We also investigated the exposure time in this paper. The result shows that there was a discontinuous expansion of the damaged area when the exposure time comes up to 2,000 ms. We can minimize the damage by keeping the exposure time less than 2,000 ms by, for example, feedback control based on detection of brighter plasma emission at the surface of the finger in contact with aerial laser plasma.

## 8 Conclusion

In this paper, we introduced a system for rendering volumetric graphics in air using a femtosecond laser. Aerial laser-induced plasma emits light without interaction with any physical matter, while one advantage of the femtosecond-laser display system is that it is safer than a system using a nanosecond laser.

There are two methods for rendering graphics in air with a femtosecond laser: Holograms by spatial light modulation technology and the scanning of a laser beam by a galvano mirror. The hologram size and workspace of the current system have maximum values of $1 \cdot cm^2$ and $5 \cdot cm^3$, respectively. Although these demonstrated sizes are currently too small to be used in the applications shown in Figure 2, this study is the first step to discuss and design laser-based aerial volumetric displays. These sizes are scalable depending on the optical devices and setup.

This paper reports the details of the theoretical principles, system setup, and experimental evaluations, and also discusses scalability, limitations, and applications. Although we focus on laser-induced plasma, the same considerations can be applied to other emission techniques such as fluorescence and cavitation.